\documentclass{elsart}
\usepackage{amsmath,amssymb,graphicx,bm}

\newcommand{\be}{\begin{equation}}
\newcommand{\ee}{\end{equation}}
\newcommand{\bea}{\begin{eqnarray}}
\newcommand{\eea}{\end{eqnarray}}
\newcommand{\ba}{\begin{align}}
\newcommand{\ea}{\end{align}}
\newcommand{\lm}{\Lambda}
\newcommand{\kf}{k_{\rm F}}

\newcommand{\vlowk}{V_{{\rm low}\,k}}

\newcommand{\tlowk}{T_{{\rm low}\,k}}

\newcommand{\vnn}{V_{\rm NN}}
\newcommand{\tnn}{T_{\rm NN}}

\newcommand{\fm}{\, \text{fm}}
\newcommand{\fmi}{\, \text{fm}^{-1}}
\newcommand{\mev}{\, \text{MeV}}

\newcommand{\ddlamfsq}{\frac{d}{d\Lambda}[f^2(p)]}
\newcommand{\openone}{\leavevmode\hbox{\small1\normalsize\kern-.33em1}}

\begin{document}

\begin{frontmatter}

\title{Dependence of the BCS $^1$S$_0$ superfluid \\
pairing gap on nuclear interactions}

\author{K. Hebeler}$^1$,
\ead{k.hebeler@gsi.de}
\author{A. Schwenk}$^{2,3}$ and
\ead{\\schwenk@triumf.ca}
\author{B. Friman}$^1$
\ead{b.friman@gsi.de}
\address{$^1$GSI, Planckstrasse 1, D-64291 Darmstadt, Germany \\
$^2$TRIUMF, 4004 Wesbrook Mall, Vancouver, BC, Canada, V6T 2A3 \\
$^3$Department of Physics,
University of Washington, Seattle, WA 98195-1560}


\begin{abstract}
We study in detail the dependence of the $^1$S$_0$ superfluid pairing
gap on nuclear interactions and on charge-independence breaking at the
BCS level. Starting from chiral effective-field theory and conventional
nucleon-nucleon (NN) interactions, we use the renormalization group
to generate low-momentum interactions $\vlowk$
with sharp and smooth regulators.
The resulting BCS gaps are well constrained by the NN
scattering phase shifts, and the cutoff dependence is very weak
for sharp or sufficiently narrow smooth regulators
with cutoffs $\lm > 1.6 \fmi$. It is therefore likely that
the effect of three-nucleon interactions on $^1$S$_0$ superfluidity
is small at the BCS level. The charge dependence of nuclear
interactions has a $10 \%$ effect on the pairing gap.
\end{abstract}

\end{frontmatter}

\section{Introduction}

Superfluidity plays a central role in strongly-interacting
many-body systems ranging from nuclei, halo nuclei and
neutron stars to cold atoms: The isospin dependence of nuclear
pairing gaps shows striking trends over a range of
isotopes~\cite{Litvinov}, the $\beta$ decay of the two-neutron halo
in $^{11}$Li is suppressed due to pairing~\cite{Sarazin} similar to
neutrino emission in neutron star cooling~\cite{Yakovlev}, and
resonant Fermi gases exhibit vortices~\cite{Zwierlein} and
superfluid characteristics in thermodynamic~\cite{Thomas} and
spectroscopic properties~\cite{Grimm}.

For relative momenta $k \lesssim 2 \fmi$,
nucleon-nucleon (NN)
interactions are well constrained by the existing scattering
data~\cite{VlowkReport}. The model dependence for larger
momenta shows up prominently, for instance, in the $^3$P$_2$
superfluid pairing gaps for Fermi momenta $\kf > 2
\fmi$~\cite{Baldo}. However, some uncertainty remains concerning
a possible dependence of the $^1$S$_0$ pairing gap on the input
NN interaction in low-density neutron matter ($\kf < 1.6
\fmi$). In this letter, we clarify this point
and explore the dependence of $^1$S$_0$ superfluidity
on nuclear interactions at the BCS level in detail.
We find that the BCS
gap is well constrained by the NN phase shifts. Therefore,
any uncertainties are due to polarization (induced interaction),
dispersion and three-nucleon (3N) interaction effects.

In addition to chiral effective-field theory (EFT) and
conventional NN interactions, we use the
renormalization group (RG) to evolve nuclear interactions
to a lower resolution scale. The resulting class of
low-momentum interactions $\vlowk$~\cite{VlowkReport,VlowkPLB,Vlowksmooth},
which is defined by a regulator with a variable
cutoff $\lm$, reproduces the NN scattering phase shifts
for momenta below $\lm$. We find that the cutoff dependence of
the $^1$S$_0$ BCS gap is very weak for sharp or sufficiently
narrow smooth regulators with $\lm > 1.6 \fmi$. A comparison
with the cutoff dependence found in $A=3,4$ nuclei
(``Tjon-line'')~\cite{Vlowk3N} and in nuclear
matter~\cite{Vlowknm}, when 3N interactions are neglected,
suggests that 3N interaction effects
on $^1$S$_0$ superfluidity are small at the BCS level.

Contact or separable pairing interactions can be implemented
directly in current density-functional calculations. For
low-momentum interactions, the weak-coupling approximation
with a density-dependent contact interaction is reliable
(see Ref.~\cite{RGnm}), and a separable approximation is
efficient~\cite{VlowkWeinberg}.
Therefore, low-momentum interactions offer the possibility
for a consistent treatment of the particle-hole and pairing
channels in density-functional theory. Moreover, it is straightforward
to adapt the RG to microscopically derive the renormalized
pairing interaction introduced in the optimal regularization
scheme of Bulgac~\cite{Bulgac}.

This letter is organized as follows. In Sect.~\ref{sect:formres},
we present the formalism for $^1$S$_0$ superfluidity at the
BCS level and our results for the dependence of the pairing
gap on nuclear interactions and for the effects of
charge-independence breaking.
We conclude in Sect.~\ref{sect:concl}.
We emphasize that our work should be considered as a
theoretical benchmark and not as a prediction of the
superfluid pairing gap, since we do not include
contributions beyond the BCS level.

\section{Formalism and results at the BCS level}
\label{sect:formres}

In the BCS approximation, the $^1$S$_0$ superfluid gap $\Delta(k)$
is obtained by solving the gap equation with a free spectrum,
$\varepsilon(p) = p^2/2$ (in units $c=\hbar=m=1$, with $m$ the
nucleon mass),
\be
\Delta(k) = - \frac{1}{\pi} \int dp \, p^2 \: \frac{V(k,p) \,
\Delta(p)}{\sqrt{\xi^2(p) + \Delta^2(p)}} \,,
\label{gapeq}
\ee
where $V(k,k')$ is the free NN interaction, $\xi(p)
\equiv \varepsilon(p) - \mu$, and for a free spectrum the chemical
potential is given by $\mu = \kf^2/2$. The
Lippmann-Schwinger equation for the scattering $T$ matrix in the
same channel reads
\be
T(k,k';E) = V(k,k') + \frac{2}{\pi} \int dp \, p^2 \:
\frac{V(k,p) \, T(p,k';E)}{E - p^2} \,.
\label{Teq}
\ee
Here and in the following, principal value integrals are implied.
The homogeneous gap equation can be understood
as an equation for the residue of the pole of the $T$ matrix,
$T(k,k';E) \to \Delta(k) \Delta(k')/[E-2\mu]$, for $E \to 2\mu$,
when the two-nucleon propagator in Eq.~(\ref{Teq}) is replaced
by the corresponding self-consistent Nambu-Gorkov propagator
$2\mu-p^2 \to - 2 \sqrt{\xi^2(p) + \Delta^2(p)}$. This includes
the propagation of back-to-back
particle-particle, $(1-n_p)(1-n_p)$, and
hole-hole modes, $- n_p \, n_p$,
where $n_p$ denotes the Fermi-Dirac distribution.

For conventional large-cutoff and chiral EFT potentials, $V(k,k')$
includes regulating functions that render the integral convergent.
These are of exponential form $\exp[-(k^2/\lm^2)^3]$ with $\lm
=450-600 \mev$ in the current chiral EFT interactions
at N$^3$LO~\cite{N3LO,N3LOEGM},
and phenomenological functions that imply large (few GeV)
cutoffs in conventional NN potential models.

For the RG evolution to low-momentum
interactions $\vlowk$, we define a reduced interaction $v(k,k')$ with
$\vlowk(k,k') = f(k) \, v(k,k') \, f(k')$, where $f(k)$ denotes a
sharp or smooth regulator. Starting from an NN interaction
with a large cutoff $v(k,k') = \vnn(k,k')$, we use the RG
equation~\cite{Vlowksmooth},
\begin{multline}
\frac{d}{d\lm} \, v(k',k) = \frac{1}{\pi} \int_{0}^{\infty} p^2dp \,
\biggl[\frac{v(k',p) \ddlamfsq \, t(p,k;p^2)}{p^2-k^2} \\[2mm]
+ \frac{t(k',p;p^2) \, \ddlamfsq \, v(p,k)}{p^2-k'^2} \biggr] \,,
\label{RGE}
\end{multline}
to generate low-momentum interactions with a
variable cutoff $\lm$ (both $v$ and $f$
depend explicitly on $\lm$). Here, the reduced $t$ matrix
$t(k,k';E)$ is defined by $\tlowk(k,k';E)
=f(k) \, t(k,k';E) \, f(k)$, where $\tlowk$ is the solution
to Eq.~(\ref{Teq}) with $V = \vlowk$. In this letter,
we consider a sharp cutoff $f(k)=\theta(\lm-k)$ and smooth
regulators of the exponential form $f(k)=\exp[-(k^2/\lm^2)^n]$,
where $n$ is a parameter that controls the smoothness.
The resulting $\vlowk$ is hermitian and preserves the low-momentum
fully-on-shell $\tnn$ matrix, up to factors of the regulator
function $\tlowk(k,k;k^2)=f^2(k)\,\tnn(k,k;k^2)$, as well
as the deuteron binding energy~\cite{Vlowksmooth}.

The RG equation, Eq.~(\ref{RGE}), is equivalent to a generalization
of the Lee-Suzuki method~\cite{LS1,LS2} and a subsequent Okubo
hermitization~\cite{Okubo} to smooth
cutoffs~\cite{Vlowksmooth,VlowkRG}. In the following, we will use
the projection operator formalism to construct low-momentum
interactions with a sharp cutoff and solve the RG equation for our
results obtained with a smooth regulator. We note that the RG
equation cannot be solved directly in the neutron-neutron $^1$S$_0$
channel for most of the conventional NN interactions (except for
the Nijmegen II potential), due to spurious resonances at high
($\sim
\text{GeV}$) momenta.
The freedom in the choice of the regulator $f(k)$ implies a scheme
dependence of the gap $\Delta(k) \sim f(k)$ at large momenta $k \gg
\kf$ (see Eq.~(\ref{gapeq})). We will restrict our results to the
gap on the Fermi surface $\Delta \equiv \Delta(\kf)$, where the
momenta are on-shell and the gap is scheme independent.

In the leading-order pionless EFT with sharp-cutoff regularization,
one has $V(k,k') = \theta(\lm-k) \, \theta(\lm-k')/[1/a_s - 2
\lm/\pi]$ (with scattering length $a_s$). The resulting
gap is cutoff independent for $\Delta \ll \mu$ and large cutoffs,
which follows from the gap equation with
$\Delta(k)=\theta(\lm-k) \, \Delta$,
\be
\frac{1}{a_s} - \frac{2}{\pi} \, \lm = - \frac{1}{\pi} \int\limits_0^\lm
dp \: \frac{p^2}{\sqrt{\xi^2(p) + \Delta^2}} \,.
\label{EFTgap}
\ee
For $\Delta \ll \mu$ and large $\lm$,
the integral is given by $2 \, [ -2 \kf + \lm + \kf
\ln(8\mu/\Delta)]$. The UV divergence cancels against the cutoff
dependence of the interaction in Eq.~(\ref{EFTgap}). This
leads to the standard result $\Delta = 8 \mu/e^2
\exp[\pi/(2\kf a_s)]$~\cite{PS}.

For the solution of the gap equation, we follow the method of
Khodel {\it et al.}~\cite{Khodel}: We first decompose the interaction
into a separable and a non-separable part
\be
V(k,k') = V(\kf,\kf) \, \phi(k) \, \phi(k') + W(k,k') \,,
\label{decomposeV}
\ee
where $\phi(k) \equiv V(k,\kf)/V(\kf,\kf)$ and $W(k,k')$ is a
suitably chosen function that vanishes when at least one argument
is on the Fermi surface ($k=\kf$). Then the gap equation,
Eq.~(\ref{gapeq}), can be replaced by an equivalent system of two
equations,
\bea
\phi(k) &=& \chi(k) + \frac{1}{\pi} \int dp \, p^2 \: \frac{W(k,p) \,
\chi(p)}{\sqrt{\xi^2(p) + \Delta^2 \, \chi^2(p)}} \,,
\label{gap1} \\[2mm]
0 &=& 1 + V(\kf,\kf) \: \frac{1}{\pi} \int dp \, p^2 \: \frac{\phi(p) \,
\chi(p)}{\sqrt{\xi^2(p) + \Delta^2 \, \chi^2(p)}} \,,
\label{gap2}
\eea
where $\Delta(k) \equiv \Delta \, \chi(k)$ with $\chi(\kf) = 1$.
This system has the advantage that the integrand in Eq.~(\ref{gap1})
vanishes on the Fermi surface, and consequently the function $\chi(k)$
is only weakly sensitive to changes of $\Delta(p)$ in the denominator.
Therefore, to a good approximation, Eq.~(\ref{gap1}) can be linearized.
In the first iteration, we solve Eq.~(\ref{gap1}) by inversion
using a sufficiently small constant Ansatz for $\Delta(p)$ in the
denominator. Next, we solve Eq.~(\ref{gap2}) for $\Delta$ using
Newton's method with the solution for $\chi(k)$ of the previous
step. The iteration of this procedure (where $\Delta(p)$
in the denominator of Eq.~(\ref{gap1}) is updated
at each step) leads to a
rapidly converging solution for the BCS gaps. We follow this method
for the smooth-cutoff $\vlowk$. For the sharp-cutoff case, we
have found that a solution of the gap equation, Eq.~(\ref{gapeq}),
by iteration with an initial guess
$\Delta(k) \sim \sqrt{|\kf \, \vlowk(k,k)|}$ also converges rapidly.
This Ansatz is motivated by the observation that a separable
approximation is meaningful for low-momentum
interactions~\cite{VlowkWeinberg}.

\begin{figure}[t]
\begin{center}
\includegraphics[clip=,width=8cm]{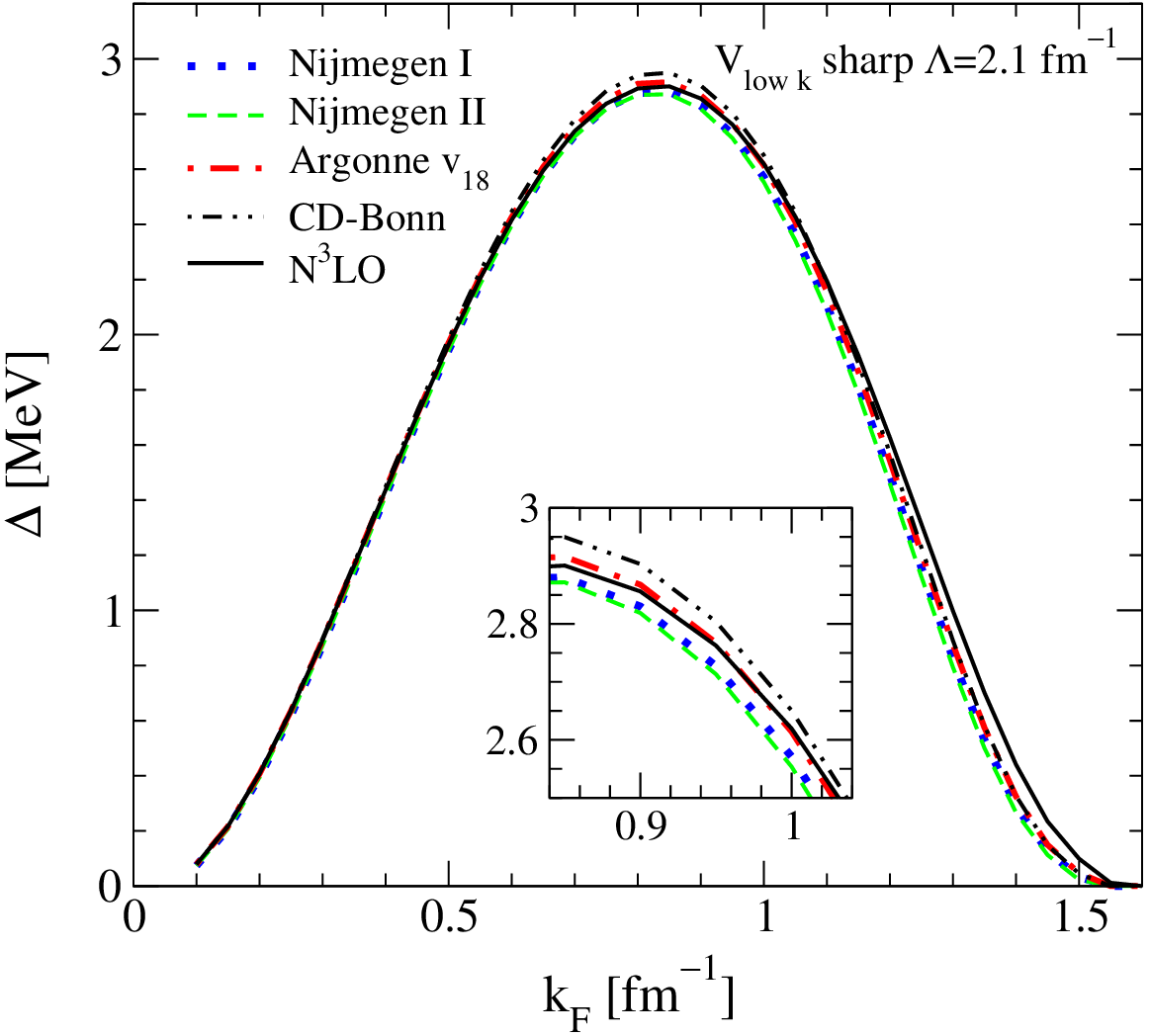}
\end{center}
\caption{The neutron-neutron $^1$S$_0$ superfluid
pairing gap on the Fermi surface $\Delta \equiv \Delta(\kf)$ versus
Fermi momentum
$\kf$ for low-momentum interactions $\vlowk$ with a sharp cutoff
$\lm=2.1 \fmi$. $\vlowk$ is derived from various charge-dependent NN
interactions~\cite{N3LO,Nijmegen,Argonne,CDBonn}.
We have verified that the results are cutoff independent from
$\lm = 1.6 \fmi$ to $\lm = 2.5 \fmi$. The inset magnifies the
small dependence on nuclear interactions near the maximum.}
\label{sharp_nn}
\end{figure}

Our results for the density dependence of the neutron-neutron
$^1$S$_0$ superfluid gap $\Delta$ are shown in Fig.~\ref{sharp_nn}.
The low-momentum interactions $\vlowk$ are derived from various
charge-dependent NN potentials~\cite{N3LO,Nijmegen,Argonne,CDBonn}
using a sharp cutoff $\lm=2.1 \fmi$. We find that
the BCS gap is almost independent of the NN interaction.
Consequently, we conclude that the $^1$S$_0$ gap is
strongly constrained by the NN scattering phase shifts. This has been
noted previously (see for example Ref.~\cite{LomSch}), but
without considering charge dependences. Moreover,
these are the first results for chiral interactions at N$^3$LO.
We use the N$^3$LO chiral potential of Ref.~\cite{N3LO}, since it
is the chiral interaction that leads to the most accuracte reproduction
of the phase shifts.

\begin{figure}[t]
\begin{center}
\includegraphics[clip=,width=8cm]{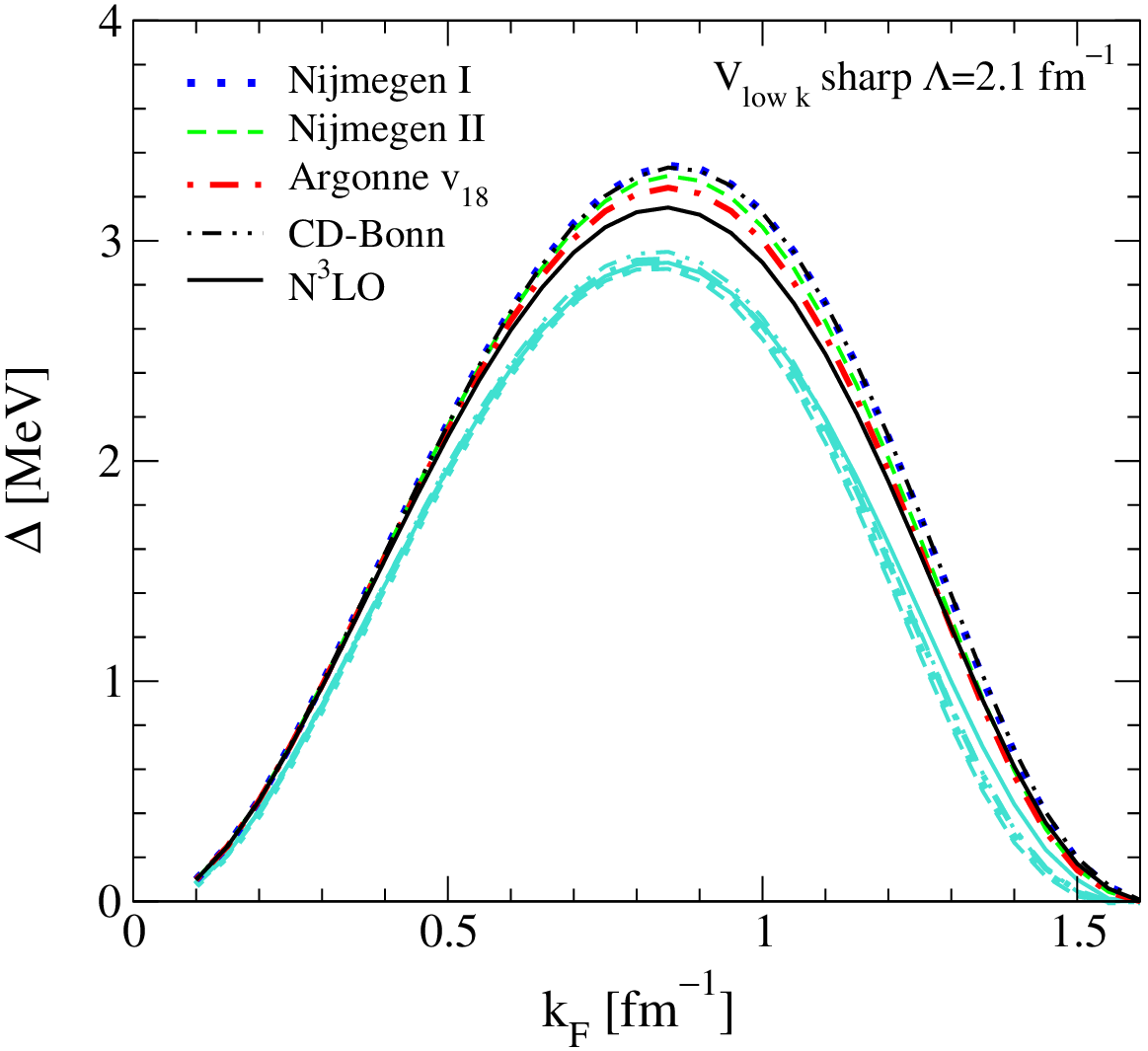}
\end{center}
\caption{The charge dependence of the $^1$S$_0$ superfluid pairing
gap $\Delta$
versus $\kf$. The lines indicated in the legend are the neutron-proton
gaps, whereas the grey lines show the neutron-neutron gaps from
Fig.~\ref{sharp_nn}. For further details, see the caption
of Fig.~\ref{sharp_nn}. We have also verified that the neutron-proton
gaps are cutoff independent over the same sharp-cutoff range.}
\label{sharp_np}
\end{figure}

The maximal gap at
the BCS level is $\Delta \approx 2.9-3.0 \mev$ for $\kf \approx 0.8-0.9 \fmi$.
The small deviation of the N$^3$LO gap from the band
at higher densities in Fig.~\ref{sharp_nn} is consistent
with the slightly more attractive $^1$S$_0$ phase shifts at the
corresponding
energies (compare, for example, the phase shifts of the
CD-Bonn~\cite{CDBonn} and N$^3$LO potentials). We find that the
gaps are cutoff independent over the range considered here, $\lm =
1.6 \fmi$ to $\lm = 2.5 \fmi$. This result
is consistent with the findings of Kaiser {\it et al.} that the
cutoff dependence is substantially reduced for chiral EFT interactions
when going from NLO to N$^2$LO~\cite{Kaisergaps}, since the latter leads
to a better description of the NN scattering phase shifts. In addition,
the BCS gaps for the ``bare'' interactions are within $2 \%$ of the
$\vlowk$ results shown in Fig.~\ref{sharp_nn} for $\kf \lesssim
1.0 \fmi$, and the difference is compatible with the spread in the
$\vlowk$ result over all densities. (This also holds for
Fig.~\ref{sharp_np}.) For completeness, we mention that
Sedrakian {\it et al.}~\cite{Sedrakian} have
solved the BCS gap equation for one low-momentum interaction ($\vlowk$
derived from Nijmegen 93~\cite{Nijmegen} with $\lm = 2.5 \fmi$), but
they did not explore the cutoff dependence.

Isospin symmetry breaking leads to small charge dependences
in nuclear interactions. As a result, the $^1$S$_0$
neutron-proton
scattering length $a_{\rm np} = - 23.768 \pm 0.006
\fm$~\cite{PWA}
is more attractive than the
neutron-neutron scattering length $a_{\rm nn} = -18.5 \pm
0.3 \fm$~\cite{Phillips} in the same channel.
This effect is dominantly due to the charge dependence of the
one pion-exchange interaction $V_\pi$. The central part of
$V_\pi$ in the neutron-proton charge-exchange channel is of the
form $- \, m_{\pi^\pm}^2/(q'^2 + m_{\pi^\pm}^2)$, where $q'$ is
the (exchange) momentum transfer. Since the charged pion is
heavier than the neutral one, $m_{\pi^\pm} = 139.57 \mev$
and $m_{\pi^0} = 134.98 \mev$, the
resulting neutron-proton interaction is more attractive.
In Fig.~\ref{sharp_np}, we show the charge dependence of
the $^1$S$_0$ superfluid pairing gap versus Fermi momentum.
The neutron-proton gaps are $\approx 0.3 \mev$
larger at maximum with a slight shift to higher densities. The
$10 \%$ effect on the pairing gaps clearly reflects the charge
dependence of nuclear interactions.

\begin{figure}[t]
\begin{center}
\includegraphics[clip=,width=8cm]{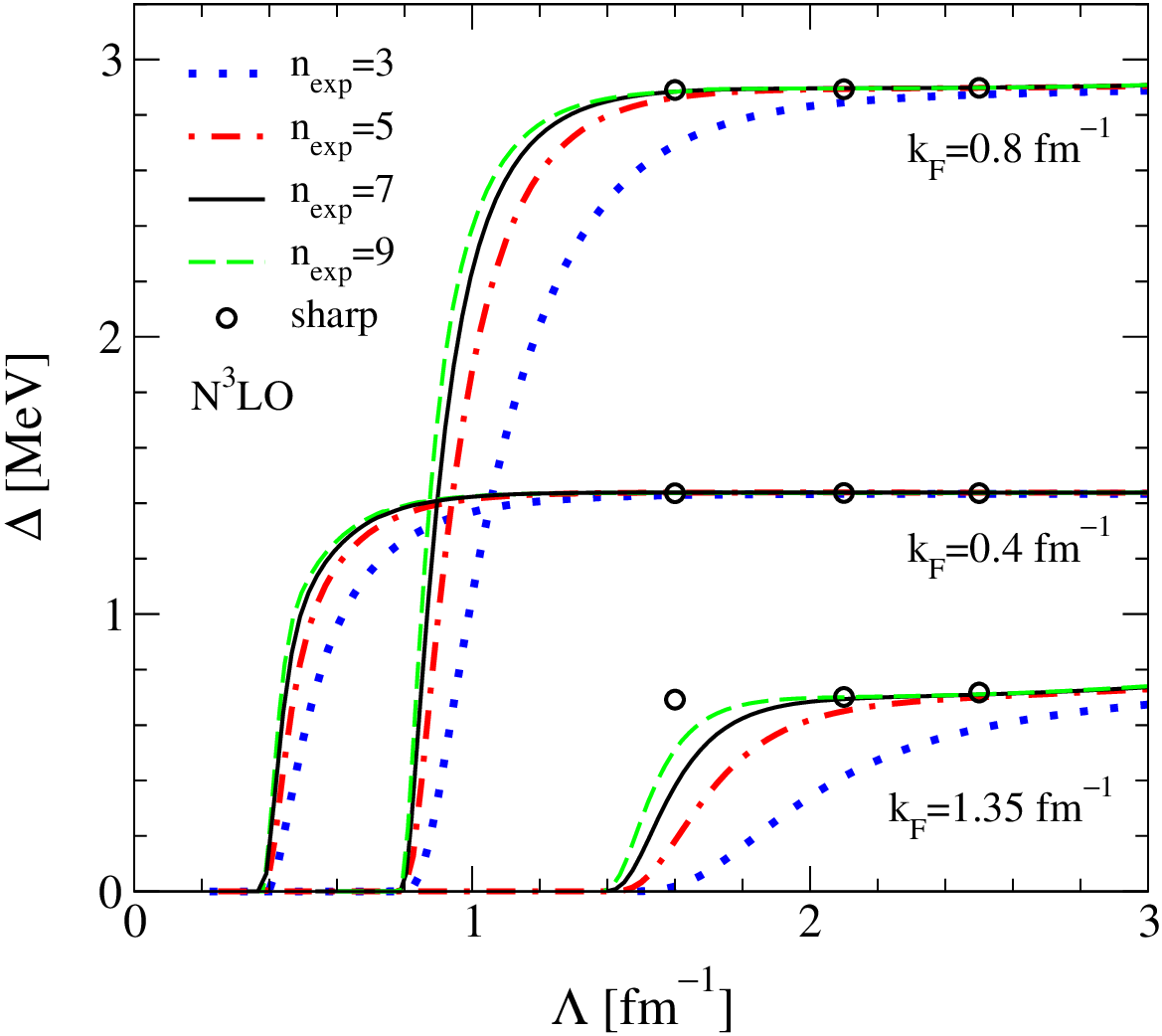}
\end{center}
\caption{The neutron-neutron $^1$S$_0$ superfluid
pairing gap $\Delta$ as a function of the cutoff $\lm$ for
three densities and different smooth exponential regulators,
as well as for a sharp cutoff. The low-momentum interactions
are derived from the N$^3$LO chiral potential of Ref.~\cite{N3LO}.}
\label{flow_nexp}
\end{figure}

\begin{figure}[t]
\begin{center}
\includegraphics[clip=,width=8cm]{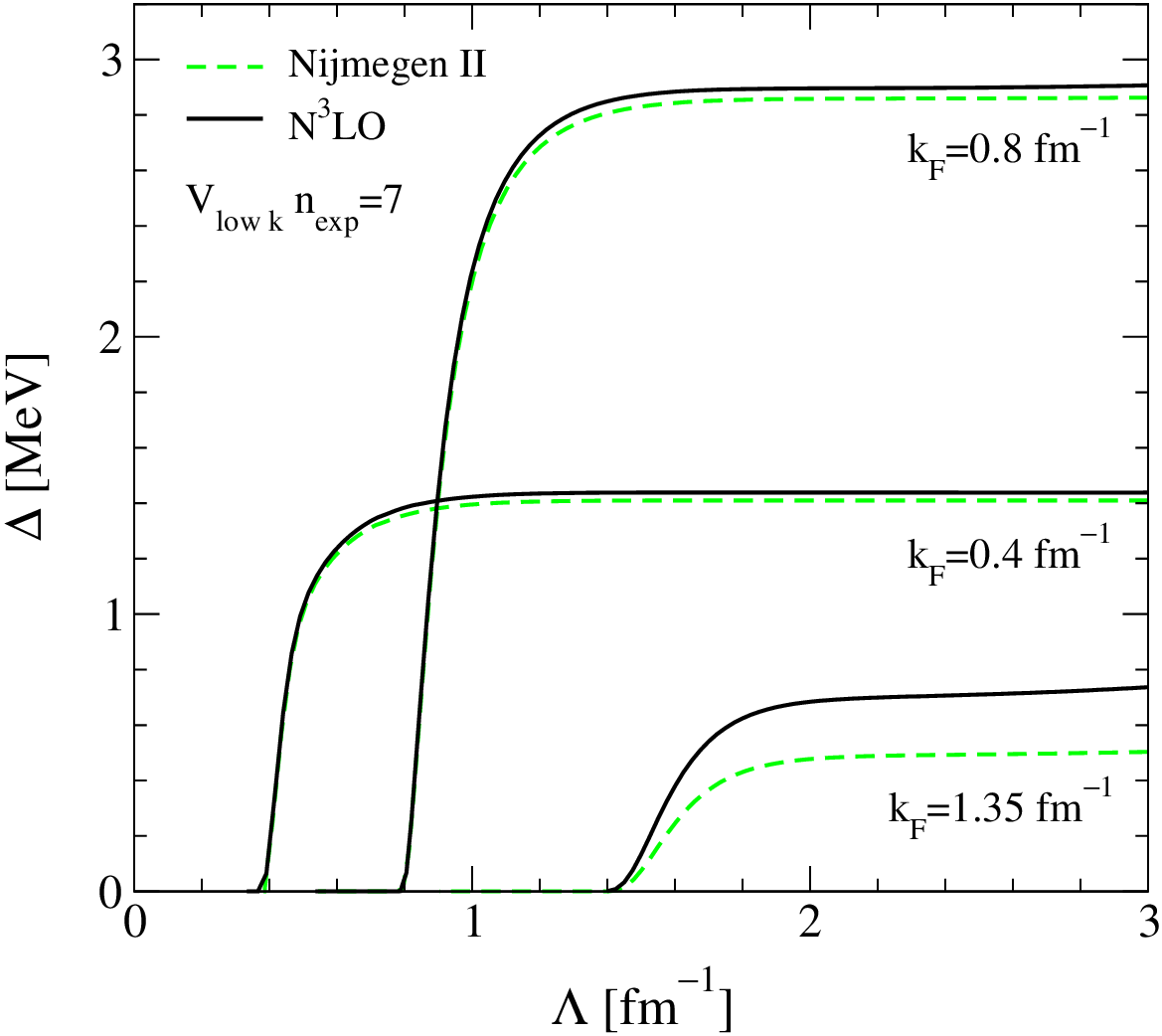}
\end{center}
\caption{The neutron-neutron $^1$S$_0$ superfluid
pairing gap $\Delta$ as a function of the cutoff $\lm$ for
three densities. The low-momentum interactions are derived
from the N$^3$LO~\cite{N3LO} and the
Nijmegen~II~\cite{Nijmegen} potential with exponential
regulator $n=7$.}
\label{flow_NN}
\end{figure}

Next, we study the dependence of the neutron-neutron $^1$S$_0$
superfluid pairing gap as a function of the cutoff starting
from the N$^3$LO chiral interaction. Our
results for three representative
densities and different smooth exponential
regulators $f(k)=\exp[-(k^2/\lm^2)^n]$, as well as for a
sharp cutoff, are shown in Fig.~\ref{flow_nexp}. As long as the
cutoff is large compared to the dominant momentum components
in the Cooper bound state, the gap depends very weakly on
the cutoff. Below this scale, which depends on the density
and the smoothness of the regulator, the strength of the
bound state decreases, since some of the momentum modes that build up
the Cooper pairs are integrated out. From Fig.~\ref{flow_nexp},
we observe that the cutoff dependence is very weak for sharp
or sufficiently narrow smooth regulators with $\lm > 1.6 \fmi$.
It can be seen that $n=3$ is too smooth, but that $n > 5$ is
sufficient. For lower densities, even lower cutoffs with $\lm
> 1.2 \kf$ are possible.

The N$^3$LO chiral interaction has a cutoff $\lm = 2.5
\fmi$ (or $500 \mev$)~\cite{N3LO} and one may suspect that the
cutoff dependence could be larger for conventional NN potentials.
In Fig.~\ref{flow_NN}, we show that this is not the case by
comparing the gaps from Fig.~\ref{flow_nexp} to results obtained
with the Nijmegen II potential~\cite{Nijmegen}, which has a large
($\sim \text{GeV}$) cutoff. The resulting cutoff dependences are 
similar and in
particular very weak for sharp or sufficiently narrow smooth
regulators with $\lm > 1.6 \fmi$. This shows that the $^1$S$_0$
superfluid pairing gap probes low-momentum physics.

\section{Conclusions}
\label{sect:concl}

In this Letter, we have presented a systematic study of the
interaction dependence of the BCS $^1$S$_0$ superfluid pairing gap.
We have shown that this gap is practically independent of the
choice of the NN interaction, and therefore well constrained by the
NN phase shifts. Furthermore, we have found only a very weak
dependence on the cutoff for low-momentum interactions $\vlowk$
with sharp or sufficiently narrow smooth regulators for $\lm > 1.6
\fmi$. For low densities, it is possible to lower the cutoff
further to $\lm > 1.2 \kf$. We also find that the pairing gap
clearly reflects the charge dependence of NN interactions.
Neutron-neutron and neutron-proton $^1$S$_0$ are not carefully
distinguished in previous work.
We conclude that the uncertainties in $^1$S$_0$ superfluidity are
due to an approximate treatment of induced interactions and
dispersion effects, which go beyond the BCS level, as well as due
to 3N interactions.

The weak cutoff dependence indicates that, in the $^1$S$_0$
channel, the contribution of 3N interactions is small at the BCS
level. We note that the $2\pi$-exchange part (``$c_i$-terms'') of
the corresponding low-momentum 3N interactions~\cite{Vlowk3N} is
cutoff independent up to the regulator functions. The latter lead
to a cutoff dependence as the density increases (see, for example,
the Hartree-Fock results in Ref.~\cite{Vlowknm}). We also emphasize
that 3N interactions will contribute differently to $^1$S$_0$
superfluidity in pure neutron matter compared to symmetric nuclear
matter. Additional insights will come from an investigation of the
cutoff dependence of $^3$S$_1$--$^3$D$_1$ superfluidity, where 3N
interactions may play an important role. Work in this direction is
in progress~\cite{3SD1}.

Low-momentum interactions, via weak-coupling or separable
approximations, can be implemented directly in current
density-functional calculations. Furthermore, it is straightforward
to adapt the RG used here to a microscopic derivation of the
optimal pairing interaction of Ref.~\cite{Bulgac}.

Finally, we emphasize that our results are obtained at the BCS
level and do not include polarization and self-energy effects.
Therefore, our work should be considered as a theoretical benchmark
and not as a prediction of the superfluid pairing gap.
Recently, polarization effects on the pairing gap have been 
studied in an RG approach to the many-body problem, starting 
from low-momentum NN interactions~\cite{RGnm}. Is is found that
polarization effects lead to a suppression to a maximal gap 
$\Delta \approx 0.8 \mev$, in qualitative agreement with
the earlier work of Wambach {\it et al.}~\cite{WAP}. The RG
approach is nonperturbative and includes long-range particle-hole
induced interactions and the dominant self-energy effects. Tensor
induced interactions are not included in Ref.~\cite{RGnm} because
their effects on $^1$S$_0$ superfluidity are expected to be small
at very low densities, and because the strength of the tensor force
is coupled to the corresponding 3N
interaction~\cite{Vlowksmooth,Vlowk3N}. In addition, there are
recent finite-particle-number, fixed-node AFD Monte Carlo 
calculations~\cite{AFDMC}, which in principle include olarization 
effects, but result in a maximal gap $\Delta \approx 2.5 \mev$.
However, in these calculations larger particle numbers may be 
required to capture long-range polarization effects.

\begin{ack}
We thank Scott Bogner, Dick Furnstahl and Philipp Reuter
for useful discussions. AS thanks the GSI Theory Group
for the warm hospitality.
This work was supported in part by the Helmholtz Association
Virtual Institute VH-VI-041, the Natural Sciences and Engineering
Research Council of Canada (NSERC) and the US Department of Energy
under Grant No.~DE--FG02--97ER41014. TRIUMF receives federal
funding via a contribution agreement through the National Research
Council of Canada.
\end{ack}


\end{document}